\documentclass[twocolumn]{article}
\usepackage{graphicx}
\usepackage{pdfpages}
\usepackage{caption}
\usepackage{abstract}
\usepackage[natbib=true, style=phys, sorting=none]{biblatex}
\usepackage{geometry}
\usepackage{authblk}
\author[1]{Cara-Lena Nies}
\author[1]{Thomas Sheerin}
\author[1,2,*]{Stefan Schulz}
\affil[1]{Tyndall National Institute, University College Cork, Lee Maltings, Dyke Parade, Cork, T12 R5CP, Ireland}
\affil[2]{School of Physics, University College Cork, College Road, Cork, Ireland}
\affil[*]{Corresponding author: stefan.schulz@tyndall.ie}
\title{Electronic and optical properties of boron containing GaN alloys: The role boron atom clustering}
\date{}
\begin{document}
\onecolumn  
\maketitle
    
\begin{abstract}
Boron (B) containing III-nitride materials, such as wurtzite (B,Ga)N alloys, have recently attracted significant interest to tailor the electronic and optical properties of optoelectronic devices operating in the visible and ultraviolet spectral range. However, the growth of high quality samples is challenging and B atom clustering is often observed in (B,Ga)N alloys. To date, fundamental understanding of the impact of such clustering on electronic and optical properties of these alloys is sparse. In this work we employ density functional theory (DFT) in the framework of the meta generalized gradient approximation (modified Becke Johnson (mBJ) functional) to provide insight into this question. We use mBJ DFT calculations, benchmarked against state-of-the-art hybrid functional DFT, on (B,Ga)N alloys in the experimentally relevant B content range of up to 7.4\%. Our results reveal that B atom clustering can lead to a strong \emph{reduction} in the bandgap of such an alloy, in contrast to alloy configurations where B atoms are not forming clusters, thus not sharing nitrogen (N) atoms. We find that the reduction in bandgap is linked mainly to carrier localization effects in the valence band, which stem from local strain and polarization field effects. However, our study also reveals that the alloy microstructure of a B atom cluster plays an important role: B atom chains along the wurtzite $c$ axis impact the electronic structure far less strongly when compared to a chain formed within the $c$-plane. This effect is again linked to local polarization field effects and the orbital character of the involved valence states in wurtzite BN and GaN. Overall, our calculations show that controlling the alloy microstructure of (B,Ga)N alloys is of central importance when it comes to utilizing these systems in future optoelectronic devices with improved efficiencies.
\end{abstract}

\twocolumn

The direct bandgap III-nitride semiconductor family (AlN, GaN and InN) and their connected alloys, (Al,Ga,In)N, have attracted great interest for achieving energy efficient optoelectronic devices such as light emitting diodes (LEDs)~\cite{humphreys08,amano2020}. In principle, by tailoring the alloy content in the active region of a device, the emission wavelength can range from the deep ultraviolet (UV) to the red~\cite{amano2020, tian22, mukai99}. However, both (deep) UV emitters utilizing (Al,Ga)N~\cite{hirayama15, shatalov12, saifaddin20, ryu13} and red emitters based on (In,Ga)N~\cite{hwang14, dussaigne20, chan21} suffer from low quantum efficiencies~\cite{kioupakis11, saifaddin20, ryu13, kudrawiec20}. Several factors contribute to these low efficiencies in the short and long wavelength window, including the large lattice mismatch in heterostructures, lattice mistmatch with the underlying substrate as well as fundamental changes in the (valence) band structure~\cite{ngo15}. To address these challenges, boron (B) containing nitride alloys have garnered significant interest in recent years. The aim is to utilise the smaller lattice parameter of wurtzite (wz) BN in comparison to (Al, Ga, In)N to create quantum-well (QW) structures with reduced lattice mismatch and to control the valence band structure~\cite{williams17, gunning17, williams19, kudrawiec20, mickevivcius19}. Ideally, controlling the strain in the system should reduce built-in polarization fields but also lattice mismatch related defect densities.

However, growth of high quality B containing III-N alloys and heterostructures presents a serious challenge. There are two main reasons for this: (i) B atoms are dissimilar from other group III elements in terms of size and electronegativity and (ii) the ground state crystal structure of BN is hexagonal, while AlN, GaN and InN preferentially crystallize in the wz phase. Due to this and the low miscibility of BN with (Al,Ga,In)N~\cite{kudrawiec20}, only small amounts ($<$10\%) of B can be incorporated into e.g. wz (B,Ga)N alloys~\cite{gunning17, ebara19, mickevivcius19, cramer17, gautier11}, as the material undergoes a spinodal phase-degradation at higher BN contents~\cite{gunning17}. While high-quality, crystalline random alloys of (B,Ga)N with a B content around 3\% have been reported~\cite{cramer17}, experimental data indicates a tendency of B atom clustering in such alloys~\cite{gautier11}. However, the impact of such clustering effects on the electronic and optical properties is largely unexplored.

To exploit the full potential benefit of B containing \mbox{III-N} alloys and heterostructures, fundamental understanding of the impact of B on the electronic and optical properties of such alloys is needed. But, in comparison to experimental studies, theoretical investigations on B containing III-N alloys are sparse and their focus is mainly on bandgap bowing parameters, miscibility limits or the direct-indirect bandgap crossover~\cite{turiansky19, shen21, shen17, williams17, williams19} when assuming random alloys. 

Clustering of B atoms in III-N alloys has been largely overlooked as bigger simulation cells are required to address experimentally relevant low B concentrations and to capture potential carrier localization effects. These larger cells often present huge computational challenges for state-of-the-art first principle methods, such as Heyd, Scuseria and Ernzerhof (HSE) hybrid functional density functional theory (HSE-DFT)~\cite{HSE,HSE06}. For example, Shen \emph{et al.}~\cite{shen17} and Turiansky \emph{et al.}~\cite{turiansky19} study the bandgap bowing in (B,Al)N and (B,Ga)N alloys, respectively, using HSE-DFT. They account for alloy fluctuations by creating ten random alloy configurations for different B concentrations in a 72 atom supercell, which allows a statistical approach to account for the changes in material properties caused by different alloy distributions. Williams and Kioupakis~\cite{williams19} study quaternary (B,Al,Ga)N alloys using special quasi random structures (SQS), to provide insight into the impact of random alloy fluctuations on the electronic structure. While their work provides a general understanding of the material properties, it does not shed light on deviations (e.g. clustering) from a random alloy configuration.

\begin{figure*}[t!]
    \centering
    \includegraphics[width=\textwidth,keepaspectratio]{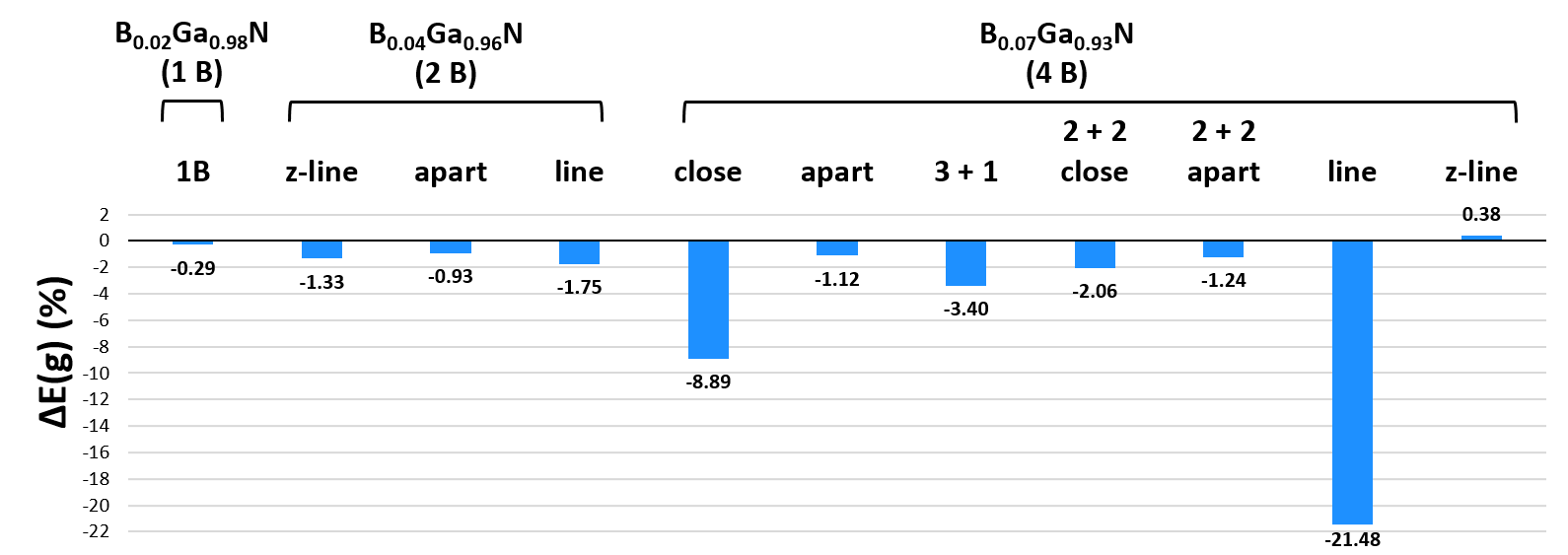}
    \caption{Percentage change of the bandgap of B$_x$Ga$_{1-x}$N alloys with respect to GaN for different atomic configurations. As the percentage change was calculated with respect to the bandgap of GaN, a negative change indicates a \emph{reduction} in the bandgap. See main text for details and explanation on the notation for the different alloy configurations.}
    \label{fig:BG-chart}
\end{figure*}

Recently, an alternative to HSE-DFT has emerged, namely the meta generalised gradient approximation (meta-GGA), which is significantly less resource intensive but produces comparative results, especially for bandgaps~\cite{borlido20}. Meta-GGAs incorporate the kinetic energy density in addition to the density gradient, which alleviates the effects of the self-interaction error~\cite{perdew81,yamamoto19} that causes bandgap underestimation in traditional GGA potentials~\cite{xiao11,perdew17,lany08}.

In this study, we apply modified Becke-Johnson (mBJ) meta-GGA to study the effects of B atom clustering on the electronic structure of wz (B,Ga)N; we also demonstrate (see Supplemental Information (SI)) that mBJ reproduce HSE-DFT band structures in the region of interest (e.g. $\Gamma$- and $K$-point) of wz-BN, AlN, GaN and InN with good accuracy. In the following, all results and discussions will focus on wz III-N, unless otherwise specified. Overall, our findings show that the alloy microstructure significantly affects the electronic and optical properties of (B,Ga)N alloys with B contents up to the here-considered 7.4\%. For instance, we find that for a fixed B content in a (B,Ga)N alloy, clustering of B atoms (B atoms sharing N atoms) can reduce the bandgap drastically compared to configurations in the absence of such clustering. This observed reduction in bandgap is mainly linked to carrier localization effects in the valence band, which we attribute to local strain and polarization field effects introduced by B atom clustering; such carrier localization effects are not observed when the B atoms are evenly distributed.

\textbf{Methods: }All calculations were carried out with DFT as implemented in the Vienna ab initio Simulation Package (VASP) v5.4~\cite{vasp}. Models and charge densities were visualised using VESTA~\cite{VESTA}. When evaluating the band structures of pristine wz GaN and BN, we used a $\Gamma$-centred Monkhorst-pack k-point mesh of 6$\times$6$\times$4, Gaussian smearing of \mbox{$\sigma$ = 0.1 eV}, and a plane wave cut-off energy of \mbox{600 eV}. The convergence criteria were $1\times10^{-4}$ eV for the energy minimisation and 0.02 eV/\AA~ for the forces in the ionic relaxations. The valence electrons for all atom types and the semicore $d$-electrons for Ga are described explicitly by expanding their wave function in a plane wave basis set, while the core electrons are treated by PAW potentials~\cite{PAW, KressePAW}. 
For our DFT investigations we used HSE functionals~\cite{HSE06} as a reference for mBJ meta-GGA~\cite{mBJ,BJ} calculations.
As mBJ meta-GGA does not allow structural relaxations, the generalized gradient approximation (GGA), as implemented in the Perdew-Burke-Ernzerhof (PBE) exchange-correlation functional~\cite{PBE}, was used for lattice and geometry relaxations. All electronic structure calculations presented below utilize mBJ meta-GGA~\cite{mBJ}. Further information on our HSE-DFT and mBJ meta-GGA benchmark are given in the SI.

To study the impact of B atom clustering on the electronic and optical properties of (B,Ga)N alloys, we used a (3$\times$3$\times$3) supercell that contains 108 atoms. One to four Ga atoms were replaced with B atoms in various configurations (see below), which corresponds to B contents ranging from $\approx$ 2 to $\approx$7\%.  
For the (B,Ga)N structures, the lattice parameters as well as internal degrees of freedom of the atoms in the simulation cell were optimised until the pressure on the system was minimised. 
We note that the lattice parameters obtained with this method deviate from the lattice parameters predicted based on a linear interpolation (Vegard's approximation) by no more than 0.7\%. Also, the introduction of B atoms into GaN did not cause any major changes in the crystal structure (no phase change).

\begin{figure}[t!]
    \centering
    \includegraphics[width=\columnwidth,keepaspectratio]{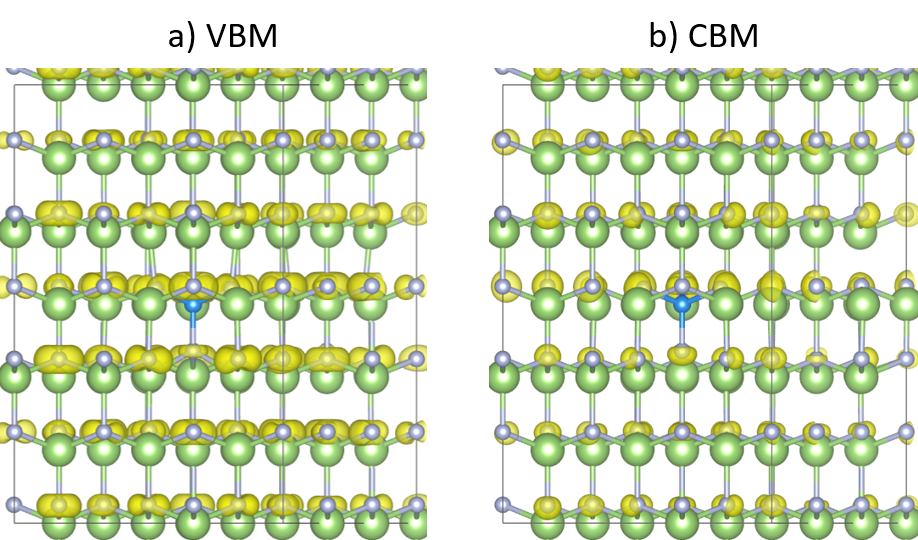}
    \caption{Band decomposed charge densities for a) the valence band maximum (VBM) and b) the conduction band minimum (CBM) for a single B atom in GaN. Ga = green spheres, B = blue spheres, N = silver spheres, charge density = yellow areas. The charge densities are plotted using a constant isosurface level of 0.0013 in VESTA. Please note that any potential gaps in the crystal structure are caused by visualising a system with periodic boundary conditions.} 
    \label{fig:1B-chgdens}
\end{figure}

\textbf{Results and Discussion: }To achieve a thorough understanding of the effect of B atom clustering on the electronic structure of (B,Ga)N alloys, we study an array of different alloy arrangements of up to four B atoms (visual representations of these configurations are included in the SI):

\begin{itemize}
    \item \textbf{Close:} 3 or 4 B atoms substituted at cation sites around a single N, i.e. in case of 4 B atoms a full tetrahedron is formed. 
    \item \textbf{Apart:} 2, 3 or 4 B atoms substituted with the maximum number of cation sites between them so that they are not sharing a N atom.
    \item \textbf{Line:} 2, 3 or 4 B atoms substituted at adjacent cation sites always connected by a N atom in the same $c$-\emph{plane}. 
    \item \textbf{z-Line:} 2 or 4 B atoms substituted vertically in a zig-zag along the $c$-\emph{axis} (i.e., one or two unit cells of BN stacked on top of each other).
    \item \textbf{2 $+$ 1:} 2 B sharing an N atom with a third B atom at the next nearest cation site in the same layer (same $c$-plane) as one of the B atoms. 
    \item \textbf{2 $+$ 2 Close:} two 2 B z-line configurations at next-nearest cation sites in the same plane.  
    \item \textbf{2 $+$ 2 Apart:} two 2 B z-line configurations at cation sites that are as far away from each other as possible.
    \item \textbf{3 $+$ 1:} a 3 B close configuration with an additional fourth B atom at the next nearest cation site in the same layer as the B atom at the top of the tetrahedron containing the 3 B atoms.
\end{itemize}

Figure~\ref{fig:BG-chart} shows the relative change in bandgap for the different alloy configurations in the studied (B,Ga)N 
alloys with respect to pristine GaN. In the following we focus mainly on the configurations with a single, 2 and 4 B atoms to discuss trends in the electronic and optical 
properties. Detailed information on the system with 3 B atoms can be found in the SI, with additional information on the bandgap changes and lattice 
constant variations for all configurations.

For the single B atom ($\approx$ 2\% B content), we find that the bandgap is not significantly changed. The charge densities, Fig.~\ref{fig:1B-chgdens}, of the conduction band minimum (CBM) and valence band maximum (VBM) of this configuration show that the electronic structure is only weakly perturbed by an isolated B atom in GaN. This finding is also observed when studying larger B contents and assuming that B atoms are not sharing N atoms ("apart" in Fig.~\ref{fig:BG-chart}): the bandgap changes by only $\approx 1$\% regardless of B content.

However, in alloy configurations where B atoms are clustering and thus sharing N atoms, the bandgap of a (B,Ga)N alloy is more significantly affected. In general, this effect increases in magnitude with increasing B content as Fig.~\ref{fig:BG-chart} reveals (see also SI for 3 B atom case). For example, the change in bandgap for 2 B atoms sharing a N atom can be $\approx -1.8$\%, while for 4 B atoms sharing the same N atom the change is $\approx-8.9$\% (for 3 B atoms the change in bandgap is $\approx -3.3$\%, see SI) when compared to pure GaN. Moreover, this bandgap \emph{reduction} also depends on the orientation of the B atoms with respect to each other. In the 2 B system, the change in bandgap is slightly larger ($\approx-1.8$\%) for B atoms in the same $c$-\emph{plane} ("line" configuration in Fig.~\ref{fig:BG-chart}) when compared to B atoms oriented along the $c$-\emph{axis} ($\approx-1.3$\%; $z$-line configuration in Fig.~\ref{fig:BG-chart}). This effect is dramatically increased in the 4 B atom case. In the "line" configuration the bandgap is changed by $\approx -21.5$\% (\emph{reduction} in bandgap compared to pure  GaN) while in the "$z$-line" arrangement the bandgap changes by $\approx +0.4$\% (slight \emph{increase} in the bandgap when compared to pure GaN). 

Overall, Fig.~\ref{fig:BG-chart} shows that, with the exception of the $z$-line configuration, the bandgap \emph{decreases} when B atoms are incorporated in  GaN; this statement also holds for the 3 B atom case (see SI). Previous calculations~\cite{turiansky19} on the bandgap of (B,Ga)N suggest that the change in bandgap is minimal in the low B content regime ($\leq 7.4$\%), and that the bandgap is expected to widen, not shrink, with increasing B content. However, we note that in Ref.~\cite{turiansky19} (i) a random alloy is assumed and (ii) the supercells studied are smaller ($3\times3\times2$, thus 72 atoms) when compared to our systems ($3\times3\times3$, thus 108 atoms). Therefore, B atom clustering as considered in our work for low B contents (up to 7\%) is most likely not resolved in Ref.~\cite{turiansky19}. Additionally, the shorter supercell size along the $c$-axis considered in Ref.~\cite{turiansky19} may provide a preference for electronic coupling of B atoms along the $c$-axis similar to our ''z-line'' configuration. All these factors may explain the differences and similarities in the bandgap evolution in (B,Ga)N alloys observed in our work and Ref.~\cite{turiansky19}.
On the other hand, recent experimental work on 500 nm (B,Ga)N epilayers grown on GaN/sapphire showed that in (B,Ga)N/GaN systems with a B content of up to 3\% a \emph{reduction} in bandgap of 0.03 to 0.04 eV occurs~\cite{mickevivcius19}. According to the findings in Ref.~\cite{mickevivcius19}, this effect stems from shifts in the valence band of the studied (B,Ga)N alloys. 
A direct comparison between our results and the experimental data however is in general not possible since further information on the alloy microstructure and the strain state of the sample would be required. Nevertheless, the experimental data gives first indications that in the low B content regime a shrinkage of the (B,Ga)N bandgap may be observed. Furthermore, previous experimental studies also indicated B atom clustering in (B,Ga)N alloys~\cite{gautier11}. Our results above show that a bandgap reduction can be promoted by B atom clustering. However, the bandgap of  wz-BN is larger than GaN, and the natural valence band offset between the two binary materials is negative (VBE in BN lower in energy when compared to GaN) and large ($\approx -1$ eV)~\cite{dreyer14}. Based on these material properties alone, a widening of the bandgap may be expected when introducing BN into GaN. From our analysis in Fig.~\ref{fig:BG-chart} we know this is not necessarily the case and it becomes clear that the electronic structure of (B,Ga)N alloys depends on the alloy microstructure. In the following we endeavour to explain why clustered B atoms have a vastly different impact on the band structure compared to distributed B atoms. To do this, it is useful to first review the fundamental band structure properties of BN and GaN.

Based on previous work~\cite{turiansky19,sheerin22} on bulk GaN and BN, conduction band states of similar symmetry ($s$-like in character) at the $\Gamma$-point in BN and GaN are energetically far apart (> 5eV) when using the valence band offset of Dreyer \emph{et al.} of \mbox{$\approx -1$ eV}~\cite{dreyer14}; furthermore the energetic separation of the conduction band edges (CBEs) in GaN ($\Gamma$-point) and BN (\mbox{$K$-point}) is  $\approx$ 2 eV for the valence band offset from Ref.~\cite{dreyer14}. As such, when alloying GaN with BN, in the low B content regime one may expect that the CBE in (B,Ga)N is only weakly perturbed by the presence of B atoms in the lattice. In comparison the offset in the valence band is smaller, ($\approx -1$ eV~\cite{dreyer14}) as mentioned above. Additionally, both GaN and BN exhibit a positive crystal field splitting energy at the $\Gamma$-point (see also SI). Therefore, in both systems and when neglecting the weak spin-orbit coupling, the valence band edge (VBE) is $p_x$- and $p_y$-like in character, and the $p_z$-like band is shifted to lower energy~\cite{turiansky19, sheerin22, punya12}. 
However, these are the properties of the ideal bulk materials. In a (B,Ga)N alloy local strain and built-in field effects will also contribute and can significantly change the electronic structure. Thus, we first analyse the charge densities of VBM and CBM obtained from DFT to 
determine whether changes to the electronic structure occur in CBM or VBM and to understand the impact of different alloy configurations on the electronic and ultimately optical properties of (B,Ga)N alloys at the atomic scale.

\begin{figure}[t!]
    \includegraphics[width=\columnwidth,keepaspectratio]{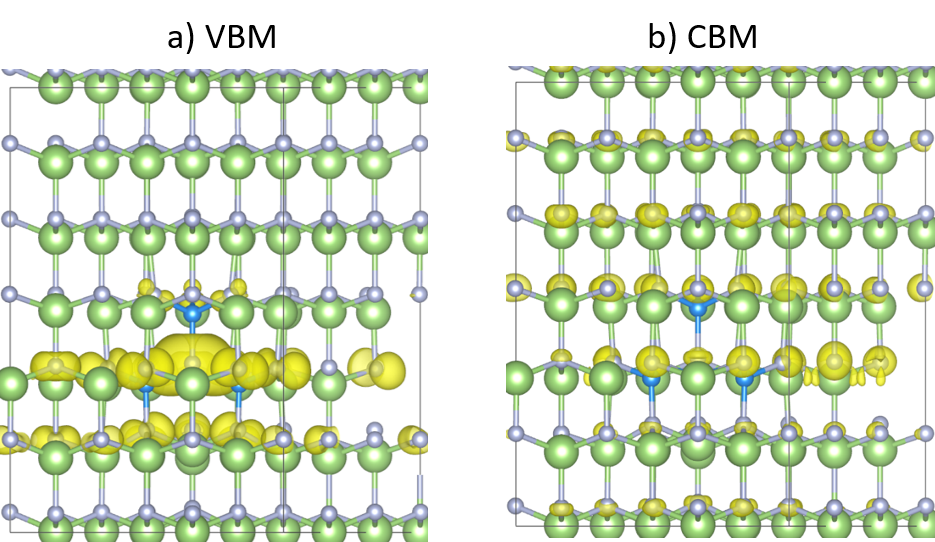}
    \caption{Band decomposed charge densities for a) the valence band maxium (VBM) and b) the conduction band minimum (CBM) for 4 B atoms sharing a N atom. For further details see caption of Fig.~\ref{fig:1B-chgdens}.}
    \label{fig:4B-close-chgdens}
\end{figure}

Figure~\ref{fig:4B-close-chgdens} displays the band decomposed charge densities for 4 B atoms sharing the same N atom (thus forming a tetrahedron around the N atom, labeled as ``close'' in Fig.~\ref{fig:BG-chart}). The charge density of the CBM indicates a delocalized state that is only slightly affected by the presence of B atoms, in line with our discussion above. However, the VBM shows strong carrier localization effects on the N atom surrounded by B atoms. The $\approx$~9\% reduction in bandgap, Fig.~\ref{fig:BG-chart}, when B atoms cluster around a N atom, can thus be traced back to carrier localization effects in the valence band.

However, given the proposed band lineup between GaN and BN~\cite{dreyer14}, such a localization effect may seem surprising at first glance. To gain further insight into the formation of such a localized valence state, and building on our DFT data, we have analyzed local strain and built-in field effects in more detail. First, we calculated a local strain tensor for the BN tetrahedron embedded in GaN; for detailed methodology and results, see SI. From our DFT simulations we find B-N bond lengths in pristine BN of $\approx 1.6$~\AA, while Ga-N bonds in bulk GaN are \mbox{$\approx$ 2.0~\AA~}. In general, in the wz crystal structure of III-N materials, bonds along the $c$-\emph{axis} tend to have a slightly different length than those in the $c$-\emph{plane}. We find that for binary systems this difference is small (0.01~\AA) but is quite pronounced in the alloys studied. Therefore, we distinguish between both types of bond lengths in the following. Our DFT calculations reveal that the B-N bonds in the 4 B tetrahedral ("close") configuration are 22\% longer when involving B atoms lying in the $c$-\emph{plane} and 9\% longer along the $c$-\emph{axis}, all compared to a relaxed BN bulk system; thus the bonds exhibit tensile strain. 
Using this information we have calculated the local strain tensor around the N atom at the center of the 4 B cluster. This strain tensor is diagonal and we find $\epsilon_{xx}\approx\epsilon_{yy}\approx 28\%$ and $\epsilon_{zz}\approx16\%$. 
Together with deformation potentials from the literature~\cite{sheerin22} and using $\mathbf{k}\cdot\mathbf{p}$ theory~\cite{schulz10} this strain tensor allows us to estimate changes in the valence band structure around the N atom. Based on this simple model, we find that $p_{x,y}$ and $p_z$-like valence states are shifted to higher energies: the $p_{x,y}$-like states will be shifted upward by 0.4 eV while the $p_z$ state is expected to be shifted by 1.5 eV with respect to their position in an unstrained BN bulk crystal. 
This may indeed indicate a localized state in the bandgap as a strain induced shift of \mbox{1.5 eV} would exceed the valence band offset of \mbox{$\approx-1$ eV}~\cite{dreyer14}. However, one may expect that the localized state is $p_z$-like in character, which is in contrast to the state predicted by our DFT calculations, see Fig.~\ref{fig:4B-close-chgdens}. 

\begin{figure}
    \centering
    \includegraphics[width=\columnwidth,keepaspectratio]{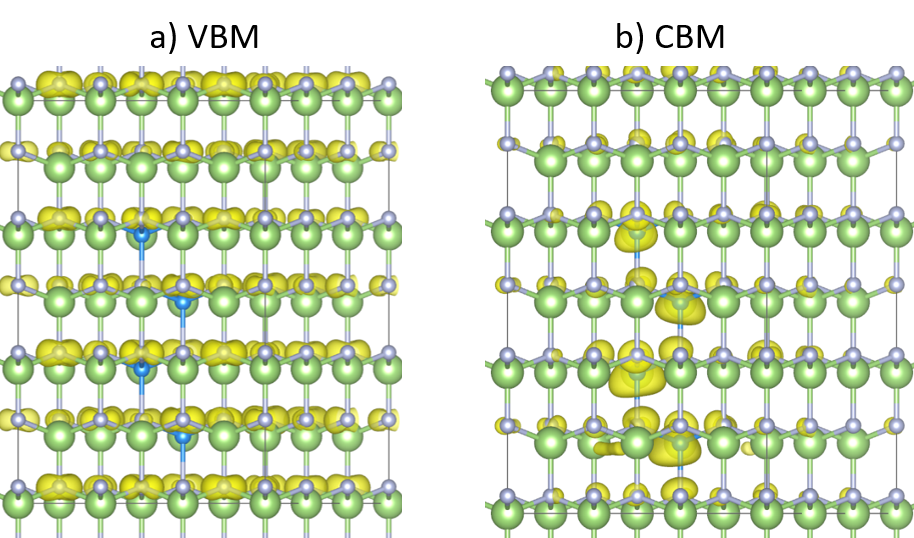}
    \caption{Band decomposed charge densities for a) the the valence band maxium (VBM) and b) the conduction band minimum (CBM) for 4 B atoms sharing a N atom, forming a line along the $c$-axis ($z$-line). For further details see caption of Fig.~\ref{fig:1B-chgdens}.}
    \label{fig:4Bzline-chgdens}
\end{figure}

In general, III-N materials are polar semiconductors which exhibit, in comparison to other III-V materials, very strong electrostatic built-in polarization fields~\cite{BeFi97}. It has already been shown that local strain effects can lead to significant local polarization fields~\cite{caro13}. These local fields may then lead to strong carrier localization effects.   
Using a simple model (see SI) to gain insight in to this question, we found that the polarization vector field in the strained BN tetrahedron has the opposite direction and is also an order of magnitude larger than the polarization vector field in bulk GaN. Thus, the discontinuity in the polarization along the $c$-\emph{axis} can lead to an electrostatic built-in field along this direction which can then result in a strong (carrier) confinement. Given that in general $p_z$-like valence states, in comparison to $p_{x,y}$-like states, have a low effective mass along $c$-\emph{axis} (see for instance Ref.~\cite{schulz10}), confinement due to local built-in fields will shift $p_z$-like valence states to lower energies when compared to $p_{x,y}$-like states. As these effects are linked to a full BN tetrahedron in GaN, these confinement effects are expected to be quite strong. Moreover, considering the high effective mass of the $p_{x,y}$-like states in BN near the $\Gamma$ point~\cite{sheerin22} one may expect carrier localization effects with $p_{x,y}$-like character at the ``interface'' between a BN tetrahedron and the GaN ``matrix''. This conforms with our DFT results shown in Fig.~\ref{fig:4B-close-chgdens}. 
Overall, our simplified analysis provides an explanation for the introduction of a localised state in the valence band, as well as its orbital character, caused by strain and polarization field effects. It also explains why the bandgap shrinks in case of such a B atom clustering. 

\begin{figure}
    \centering
    \includegraphics[width=\columnwidth,keepaspectratio]{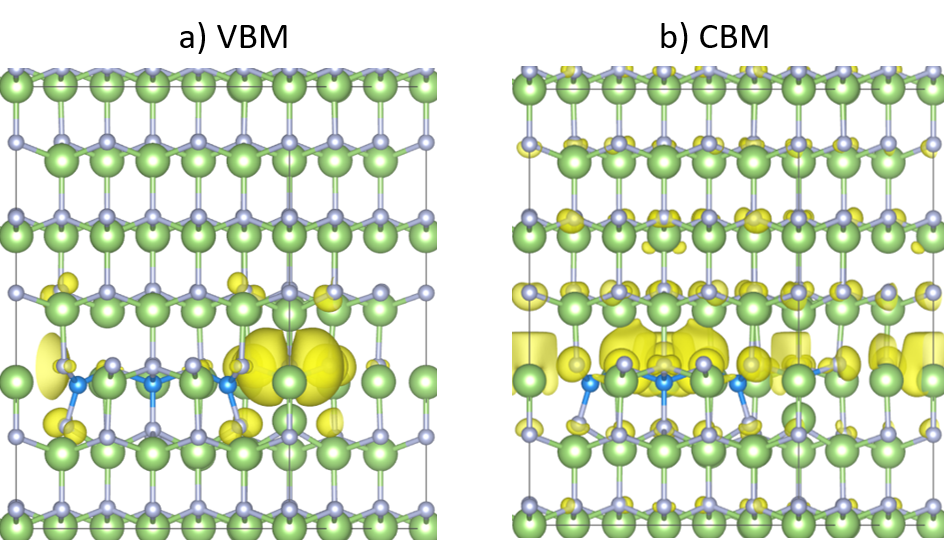}
    \caption{Band decomposed charge density for the a) the second band below the VBM, b) the band below the VBM, c) the VBM and d) the CBM of the B$_{4}$ line configuration in GaN.}
    \label{fig:4Bline-chgdens}
\end{figure}

The $p_x$- and $p_y$-like VBE character and carrier localisation facilitated by local strain and polarization effects can also explain our observation that the bandgap reduction is stronger when atoms form a line in the $c$-\emph{plane} in comparison to a line along the $c$-\emph{axis}. In the latter case one may expect that the perturbation of the VBE is negligible due to a negligible ``interface'' between BN and GaN in the $c$-\emph{plane} as well as that $p_x$- and $p_y$-like orbitals are oriented perpendicular to the stacking of the B atoms along the $c$-axis.
Indeed, Fig.~\ref{fig:4Bzline-chgdens} confirms that the charge density of the VBE is only weakly perturbed when the 4 B atoms form a ``line'' along the $c$-\emph{axis}. However, interestingly Fig.~\ref{fig:4Bzline-chgdens} indicates some weak carrier localization in the CBM. In the band structure of bulk BN~\cite{sheerin22}, there are $s$-like conduction band states at the $\Gamma$-point, that are indicative of having a high effective mass along the $c$-\emph{axis} ($\Gamma$- to $A$-direction in the first Brillouin zone). This behavior coupled with a larger spatial extent of the BN chain along the $c$-\emph{axis} may explain the observed CBM localization effect. However, further studies are required to shed light onto this effect. Nevertheless, the difference in localization characteristics (negligible localization in VBM, weak localization in CBM) may also explain why in this case the bandgap increases rather than decreases (see Fig.~\ref{fig:BG-chart}).       

Our discussion above also provides first insight into the configuration with 4 B atoms forming a line in the $c$-\emph{plane}, which presents such a drastic case in terms of the bandgap reduction: the electronic interaction between $p_x$- and $p_y$-like orbitals is enhanced by such an atomic arrangement of B atoms and the interface between BN and GaN in the $c$-plane is increased (and so is the $c$-axis oriented surface area~\cite{schulz10b}) in comparison to the ''z-line'' or 4 B atom ''close'' configuration. The expected carrier localization effect in the VBM is confirmed by our DFT calculations, as Fig.~\ref{fig:4Bline-chgdens} shows; interestingly here we also find indications of localization effects in the CBM. 
Moreover, in this alloy configuration analysis of the bond angles suggests that the B atoms are trying to form a structure resembling \emph{hexagonal} BN, the ground state crystal structure of BN. This together with our general discussions above may now explain the strong reduction in the bandgap and increased interaction in the $c$-plane of (B,Ga)N containing such a planar arrangement of B atoms.

\textbf{Conclusion: } Overall, our results show that clustering of B atoms in (B,Ga)N alloys, previously observed in experimental studies, may lead to localised states in the valence band, which can cause the bandgap to shrink in the studied B content range of $\leq 7.4$\%. These localization effects are introduced by local strain and built-in field effects. Additionally, we find that changes in the bandgap are also tightly linked to local arrangements of B atoms within a cluster: when B atoms form a line in the same $c$-\emph{plane}, this perturbs the electronic structure and thus the bandgap more significantly when compared to B atoms forming a line along the $c$-axis.
In contrast, for alloy configurations where B atoms are distributed throughout the crystal (B atoms not sharing N atoms), the bandgap only changes slightly and localization effects are negligible. 

Thus, our investigation of the impact of B atom clustering on the electronic and optical properties of (B,Ga)N alloys highlights the need for controlling the alloy microstructure. If (B,Ga)N is to be used in optoelectronic devices, a film with distributed/separated B atoms may facilitate a structure where (B,Ga)N can be used to lattice-match layers of the heterostructure to reduce strain and consequently also strain-dependent piezoelectric polarization fields. At the same time, the bandgap may only be very slightly affected. As such, \emph{if} B clustering can be avoided, this may help to improve the efficiency of e.g., UV light emitting devices without affecting the emission wavelengths of the device.

On the other hand, while the introduction of localised states could potentially help to keep carriers away from non-radiative recombination centres (Shockley-Read-Hall recombination), it may also impact radiative and non-radiative (Auger) recombination rates in a non-trivial way as seen for instance in (In,Ga)N quantum well systems~\cite{JoTe2017,McKi2022,McKi2023E,BaMc2023}. Additionally, our analysis shows that disorder, depending on the alloy atomic configuration, may significantly affect the bandgap. Therefore, designing a device of a particular emission wavelength may be challenging if the alloy microstructure is not controlled. Further experimental and theoretical studies on the electronic, optical and carrier transport properties are now required to understand their consequences for designing optoelectronic devices utilizing this new material system. 

This work received funding from the Science Foundation Ireland (Nos. 12/RC/2276 P2 and 21/FFP-A/9014).

\printbibliography



\section*{Supplementary material}

# Supporting Information: Electronic and optical properties of boron-containing GaN alloys: The role of boron atom clustering

Cara-Lena Nies[1], Thomas P. Sheerin[1], and Stefan Schulz[1,2,*]

[1]Tyndall National Institute, University College Cork, Lee Maltings, Dyke Parade, Cork, T12 R5CP, Ireland

[2]School of Physics, University College Cork, College Road, Cork, Ireland

[*]Corresponding author: stefan.schulz@tyndall.ie

# Contents

1

# S1: Comparison between HSE and mBJ meta-GGA: Computational setting, material parameters and band structures

Heyd, Scuseria and Ernzerhof (HSE) density functional theory (DFT) is currently widely employed when modelling the III-N materials AlN, GaN, InN but also BN, given that this hybrid functional approach provides an accurate description of the band gap of these materials [1]. However, as already mentioned in the main text, the computational demand of this method is quite high, which usually constrains calculations to smaller supercells when treating alloy disorder in III-N materials. On the other hand, the meta generalized gradient approximation (meta-GGA) in the framework of the modified Becke Johnson (mBJ) potential has recently been shown to also provide accurate bandgaps for semiconductor materials [2, 3]. The computational load of such meta-GGA DFT calculations is significantly reduced in comparison to a HSE-DFT treatment [4–7]. Moreover, like the exact exchange parameter, $\alpha$, in HSE-DFT functionals, where $\alpha$ controls the mixing of semi-local and Hartree-Fock exchange contributions, the mBJ meta-GGA exhibits a system dependent and adjustable constant, $c$, which can be used for instance to reproduce experimental results or HSE-DFT data. While the main aim of the present study was to gain insight into the impact of B atom clustering on the electronic and optical properties of (B,Ga)N alloys, our target was also to establish an mBJ meta-GGA based framework for III-N materials that has been benchmarked against HSE-DFT. Such an mBJ model is of benefit for the present study but also to perform future calculations on other III-N materials relevant for optoelectronic devices, e.g., (B,In,Ga,Al)N alloys at reduced computational cost but similar accuracy to full HSE-DFT calculations. Therefore, below we present results not only for wurtzite (wz) BN and GaN but also for InN and AlN.

For our comparison between HSE-DFT and mBJ meta-GGA, we proceeded as follows. Given that the bandgap is of central importance to our study, we optimised the $\alpha$ and $c$ parameters for HSE and mBJ, respectively, to reproduce the experimental bandgaps for AlN, GaN and InN, see Table S1. For all our calculations we first optimised the lattice parameters of a single unit cell of wz BN, AlN, GaN and InN until the pressure in the system was minimised. For HSE-DFT, the ionic relaxation has been performed using the HSE hybrid functional, while GGA (PBE [8]) was employed in the case of mBJ meta-GGA investigations, because the mBJ functional is a potential-only functional and does not allow for ionic relaxations [4, 9, 10]. Furthermore, and based on symmetry considerations, the positions of the anions (N atoms) were relaxed only along the $c$-axis, while all other internal degrees of freedom were constrained. This symmetry-constrained approach allows us to efficiently find the equilibrium positions of the atoms in the unit cell. We stress that for the (B,Ga)N supercell calculations, all internal degrees of freedom of all atoms in the cell are allowed to change to find the equilibrium positions of the atoms. The obtained lattice

|                    | BN                | AlN         | GaN         | InN         |
|--------------------|-------------------|-------------|-------------|-------------|
| HSE a (Å)          | 2.5321            | 3.0980      | 3.174129    | 3.5442      |
| HSE c (Å)          | 4.1868            | 4.9597      | 5.164577    | 5.7122      |
| GGA a (Å)          | 2.5546            | 3.1273      | 3.2176      | 3.5807      |
| GGA c (Å)          | 4.2244            | 5.0172      | 5.2437      | 5.7952      |
| Exp. a (Å)         | 2.55 [14–16]      | 3.110 [17]  | 3.190 [17]  | 3.544 [18]  |
| Exp. c (Å)         | 4.20 - 4.23 [14–16] | 4.980 [17] | 5.189 [17]  | 5.718 [18]  |
| $\alpha$           | 0.284382          | 0.316125    | 0.284382    | 0.240376    |
| $c$-parameter      | 1.3               | 1.6         | 1.592       | 1.315       |
| HSE Bandgap (eV)   | 6.706 ($\Gamma - K$) | 6.09     | 3.51        | 0.69        |
| mBJ Bandgap (eV)   | 6.699 ($\Gamma - K$) | 6.089    | 3.509       | 0.690       |
| Exp. Bandgap (eV)  | N/A               | 6.09 [19]   | 3.51 [20]   | 0.69 [21]   |

Table S1: Data for hybrid (HSE) DFT and meta-GGA calculations. See main text of SI for further information.

constants are shown in Table S1. In general our GGA relaxations result in sightly larger $a$- and $c$-lattice constants compared to HSE, but the values are nevertheless in good agreement with experiment, see Table S1.

We note that there is no experimental bandgap data on *wz*-BN currently available. As such, to obtain the electronic structure of wz BN in HSE-DFT, the HSE-DFT settings of GaN have been used, as GaN is the host material for the here-considered (B,Ga)N alloys. In the case of meta-GGA the $c$-parameter was optimised to reproduce the (indirect) bandgap predicted by our HSE-DFT calculations. We note that our predicted indirect bandgap of 6.7 eV between valence band maximum at the $\Gamma$-point and conduction band minimum at $K$-point is in very good agreement with HSE-DFT results in the literature [11–13]. Table S1 summarises the obtained bandgaps and corresponding optimised parameters $\alpha$ and $c$ for HSE and mBJ for each material.

While above we have mainly focused on bandgaps and optimization of all required parameters (see Table S1), below we also demonstrate that band structures obtained within mBJ meta-GGA are in good agreement with HSE-DFT band structures for the here-considered AlN, GaN, InN and BN systems. Thus, mBJ is expected to provide a robust way to determine the electronic structure of (In,Ga,Al,B)N alloys with similar accuracy to state-of-the-art HSE-DFT calculations but at a reduced computational cost. Figure S1 depicts the band structures for GaN and BN along the $k$-point path $H - K - \Gamma - A$. As Fig. S1 (a) shows, mBJ provides a very good description of the HSE data near the $\Gamma$-point, the region that is of major importance for our (B,Ga)N systems with low B contents. For BN, Fig. S1 (b), we also find very good agreement between mBJ and HSE at the valence band maximum, located at the $\Gamma$-point. We note that both mBJ and HSE predict a *positive* crystal field splitting energy of around 250 meV. Thus the valence band ordering at the $\Gamma$-point is the same in BN and GaN (both exhibit positive crystal field splitting energies), which is in contrast to AlN. Looking at the conduction band at the $\Gamma$-point, we find a fair agreement between HSE and mBJ band structure, cf. Fig. S1 (b). However, given that we are interested in low B contents and that in BN the conduction band energies at $\Gamma$ are at much higher energies when compared to

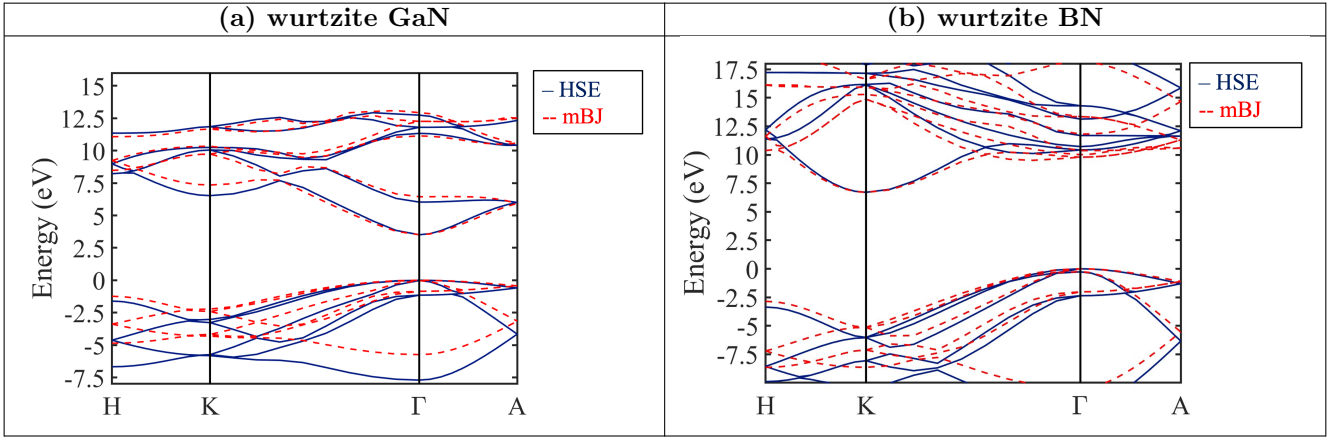

Figure S1: Electronic band structures for (a) GaN and (b) BN calculated using HSE (blue solid line) and mBJ meta-GGA (red dashed line).

GaN (even taking the band offset of -1 eV from Ref. [12] into account), any observed deviations are of secondary importance for our present studies. For the conduction band in BN at the $K$-point (conduction band minimum) mBJ and HSE agree very well, as expected from our aboved described fitting approach of the $c$-parameter.

Figure S2 displays the band structure comparison for AlN, Fig. S2 (a), and InN, Fig. S2 (b). Given that both InN and AlN are direct bandgap materials, again the band structure around the $\Gamma$-point is of key importance for the optical properties of these systems and their connected alloys. As Fig. S2 shows, for both conduction band minimum and valence band maximum at the $\Gamma$-point, mBJ and HSE band structures are in very good agreement. Only for valence states far from the valence band edge at $\Gamma$ (below -5 eV), noticeable differences are observed. However, in general for radiative recombination processes these deviations are of secondary importance.

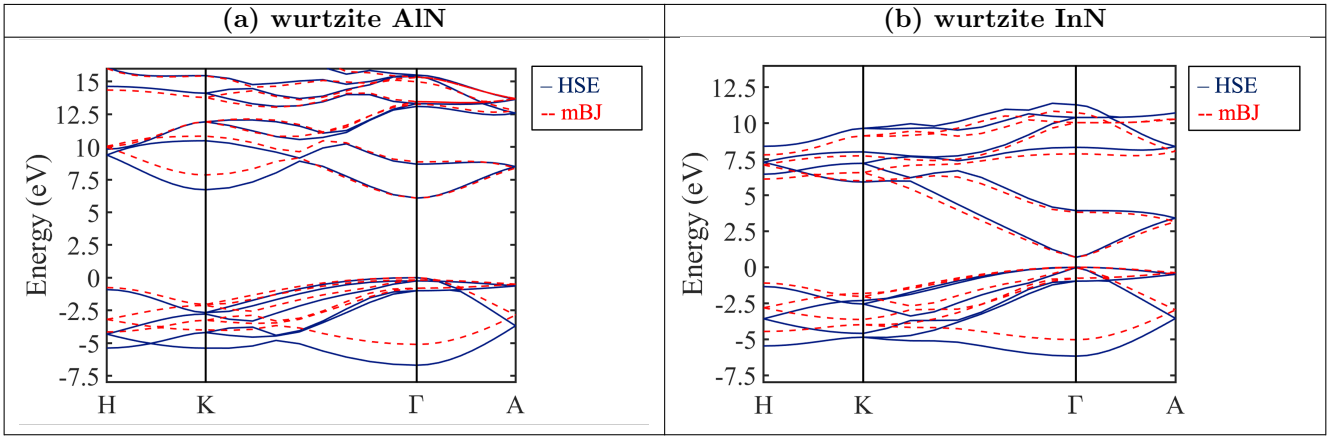

Figure S2: Electronic band structures for (a) AlN and (b) InN calculated using HSE (blue solid line) and mBJ meta-GGA (red dashed line).

4

## S2: Additional Data for (B,Ga)N geometries

In the main text of the manuscript we have discussed the relative bandgap change in (B,Ga)N alloys with 1, 2 and 4 B atoms in a 108 atom supercell for different microscopic alloy configurations. In Table S2 information on the absolute bandgap values for the different alloy contents and configurations are given. Also systems with 3 B atoms (B content of 5.6%) are reported in Table S2, further confirming that the magnitude in bandgap reduction may increase with increasing B content. Additionally, we provide details on the lattice constants of the relaxed (B,Ga)N supercells. We note that in general optimized $\alpha$ and $c$-parameters for HSE and mBJ calculations differ between the studied binary systems, see Table S1, as $\alpha$ and $c$ are here used to match the bandgap values of the different materials. In previous HSE-DFT calculations on (B,Ga)N and (B,Al)N alloys, and given that only a single, constant $\alpha$ parameter can be used in the simulation of supercells, the optimized $\alpha$ parameter for GaN or AlN was also used for BN in the case of (B,Ga)N and (B,Al)N ($\alpha_{\text{GaN}} = 0.287$ [13],$\alpha_{\text{AlN}} = 0.33$ [22]). This means that the band structure for BN, when using different $\alpha$-parameters, will change; it may also be difficult to transfer such an approach to describe the electronic structure of quaternary alloys, e.g., (B,Ga,Al)N alloys. Our optimized mBJ $c$-parameter for BN differs from the $c$-parameter in e.g., GaN. However, instead of fixing the $c$-parameter in the alloy to the value of the host material, e.g. GaN, we use here, as a first approximation, a linear interpolation between the $c$-parameters of BN and GaN based on the B content in the system. The employed $c$-parameters for the different alloy contents are given in the table below. Our approach of using a linear interpolation allows the same treatment for quaternary alloys in future studies. We note also that while it is possible to assign individual $c$-parameters to each atom type in the system, thus using optimised parameters for each element, we found that this approach produced poor agreement with HSE band structures, for instance, in terms of crystal field splitting energies and valence band ordering.

| Configuration | Doping | $c$-parameter | Lattice Parameter | | Band Gap (eV) |
| --- | --- | --- | --- | --- | --- |
| | | | $a$ (Å) | $c$ (Å) | |
| GaN | 0 % | 1.592 | 9.6527 | 15.7311 | 3.509 |
| BN | 100 % | 1.3 | 7.6638 | 12.6732 | 6.699 ($\Gamma - K$) |
| Single B | 1.9 % | 1.587 | 9.6180 | 15.6690 | 3.499 |
| $B_2$ close | 3.7 % | 1.581 | 9.5840 | 15.6095 | 3.462 |
| $B_2$ apart | 3.7 % | 1.581 | 9.5815 | 15.6123 | 3.476 |
| $B_2$ line | 3.7 % | 1.581 | 9.5678 | 15.6426 | 3.448 |
| $B_3$ close | 5.6 % | 1.576 | 9.5570 | 15.5479 | 3.393 |
| $B_3$ apart | 5.6 % | 1.576 | 9.5460 | 15.5505 | 3.466 |
| $B_3$ line | 5.6 % | 1.576 | 9.5622 | 15.5575 | 3.436 |
| $B_3$ 2 + 1 | 5.6 % | 1.576 | 9.5485 | 15.5477 | 3.453 |
| $B_4$ close | 7.4 % | 1.57 | 9.5282 | 15.4918 | 3.197 |
| $B_4$ apart | 7.4 % | 1.57 | 9.5089 | 15.4920 | 3.470 |
| $B_4$ line | 7.4 % | 1.57 | 9.5737 | 15.5098 | 2.755 |
| $B_4$ z-line | 7.4 % | 1.57 | 9.5529 | 15.5787 | 3.522 |
| $B_4$ 2 + 2 close | 7.4 % | 1.57 | 9.5152 | 15.4821 | 3.488 |
| $B_4$ 2 + 2 apart | 7.4 % | 1.57 | 9.5116 | 15.4958 | 3.492 |
| $B_4$ 3 + 1 | 7.4 % | 1.57 | 9.5185 | 15.4907 | 3.429 |

Table S2: Data for BGaN cluster geometries.

## S3: Boron Alloy Microstructures

In the following we illustrate the arrangements of B atoms in the different alloy configurations discussed in the main text and indicated in Table S2. Here we show only the B atoms in the unit cell in Fig. S3, as the B atoms are mainly located in the centre of the unit cell and this approach makes it easier to understand the position of the B atoms with respect to each other; however, we show the GaN cell for reference in Fig. S4. We note that the 4 B "line" configuration is not a perfectly straight line, as this would be impossible to achieve in a $(3 \times 3 \times 3)$ supercell. Instead, the fourth atom is located at the closest possible cation site in the plane.

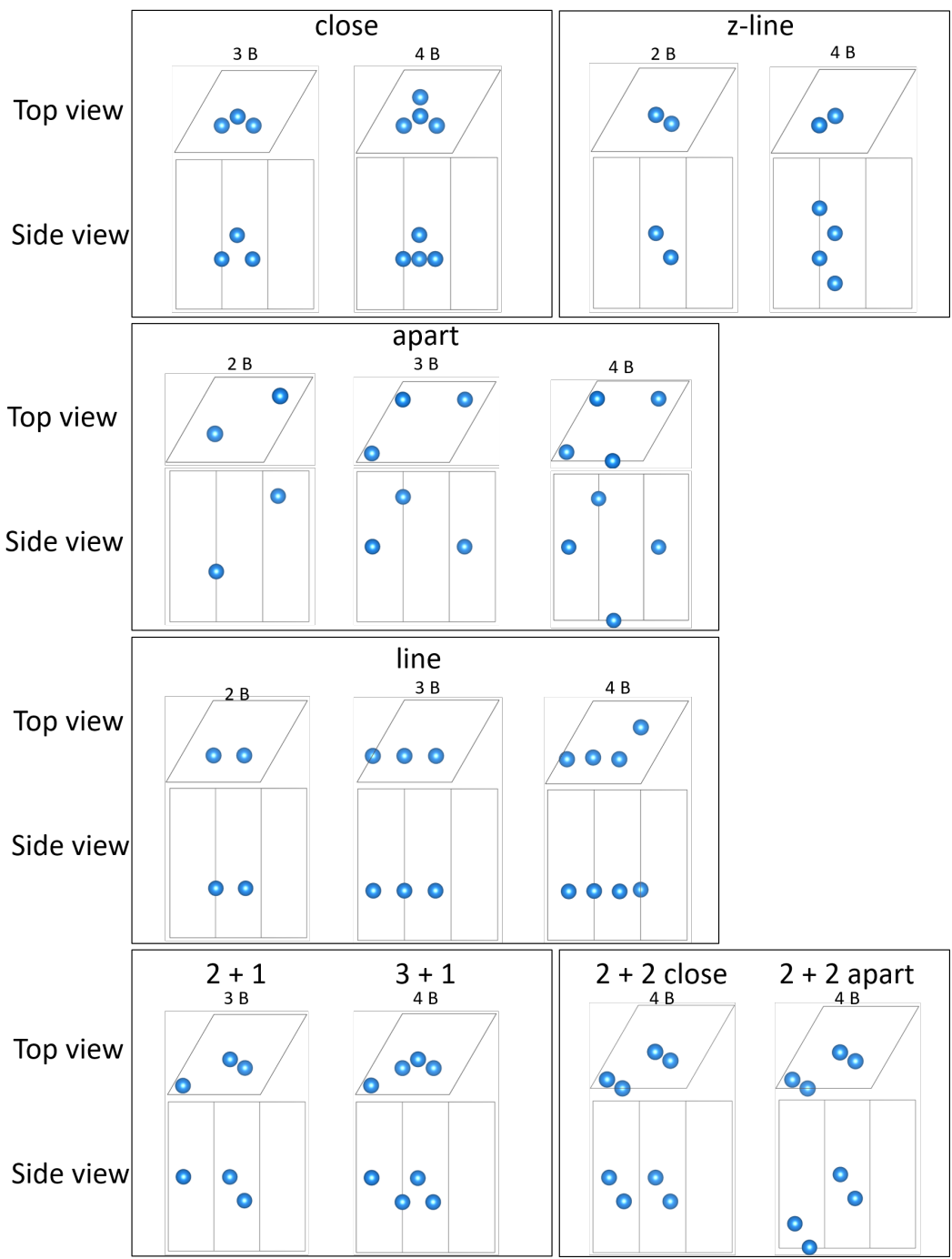

Figure S3: Location of B atoms for each configuration in the GaN unit cell box. Top view looks down at the $c$-plane and side view shows the view perpendicular to the $c$-axis.

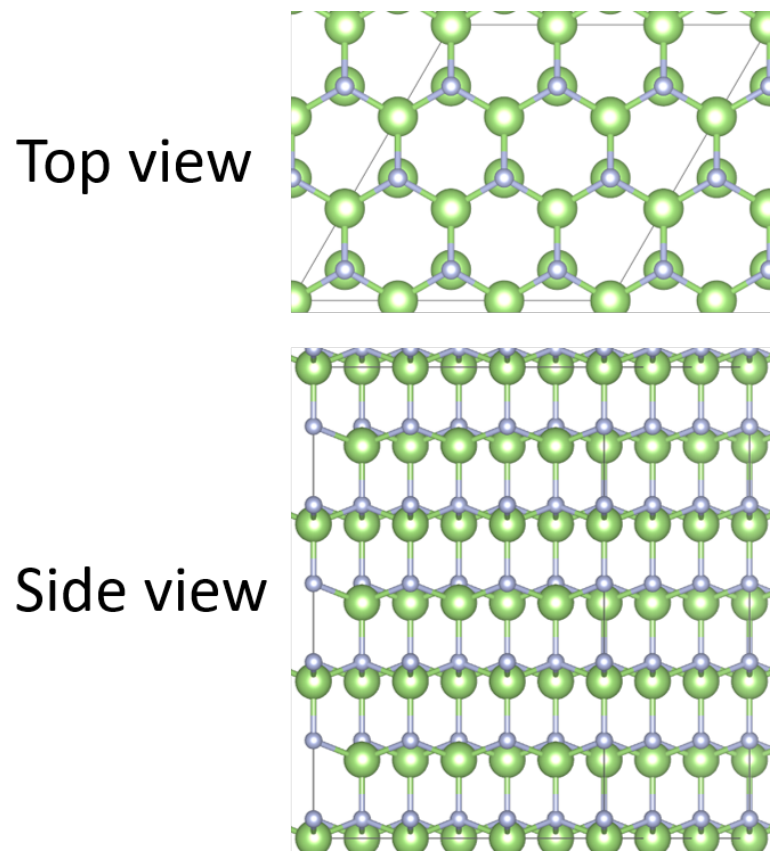

Figure S4: GaN (3 × 3 × 3) supercell as a reference point for the location of the B atoms in each configuration in Fig. S3.

## S4: Local Strain Tensor

To gain first insight into the impact that local strain effects may have on the electronic structure of (B,Ga)N alloys when B atoms cluster, we employ a blend of $\mathbf{k} \cdot \mathbf{p}$ theory and a simple model to extract a local strain tensor $\epsilon$. To obtain a local strain tensor, $\epsilon$, different approaches have already been presented in the literature [23, 24] and we follow a similar procedure. To determine $\epsilon$ at an atomic site $O$, we consider the atomic coordinates of its nearest neighbor atoms forming a tetrahedron surrounding $O$. The vectors connecting the atoms $\alpha$ and $\beta$ in the ideal (unstrained) tetrahedron are denoted by $\mathbf{R}^0_{\alpha,\beta}$, see Fig. S5. In an alloy, or in e.g., a BN tetrahedron embedded in GaN, the atoms surrounding the central atom $O$ are displaced from their equilibrium position they would occupy in a binary bulk material. Thus after the ionic relaxation in our DFT calculations the tetrahedron is distorted; the edges of the distorted tetrahedron are denoted by $\mathbf{R}_{\alpha,\beta}$, see Fig. S5. The two geometries are related to each other via:

$$\begin{pmatrix} R_{12,x} & R_{23,x} & R_{14,x} \\ R_{12,y} & R_{23,y} & R_{14,y} \\ R_{12,z} & R_{23,z} & R_{14,z} \end{pmatrix} = \begin{pmatrix} 1+\epsilon_{xx} & \epsilon_{xy} & \epsilon_{xz} \\ \epsilon_{yx} & 1+\epsilon_{yy} & \epsilon_{yz} \\ \epsilon_{zx} & \epsilon_{zy} & 1+\epsilon_{zz} \end{pmatrix} \times \begin{pmatrix} R^0_{12,x} & R^0_{23,x} & R^0_{14,x} \\ R^0_{12,y} & R^0_{23,y} & R^0_{14,y} \\ R^0_{12,z} & R^0_{23,z} & R^0_{14,z} \end{pmatrix} \quad (1)$$

The local strain tensor at the atomic site $O$ can then be obtained via matrix inversion.

We use the approach outlined above to calculate $\epsilon$ at the N site when surrounded by 4 B atoms embedded in GaN. This corresponds to the configuration labeled "close" in the 4 B case (see above and main text). We extract the Cartesian coordinates of the 4 B atoms around the central N atom in wz BN ($\mathbf{R}^0_{12}$, $\mathbf{R}^0_{23}$, $\mathbf{R}^0_{14}$) and the coordinates of the same atoms when the BN tetrahedron is embedded in GaN ($\mathbf{R}_{12}$, $\mathbf{R}_{23}$ and $\mathbf{R}_{14}$) from our DFT

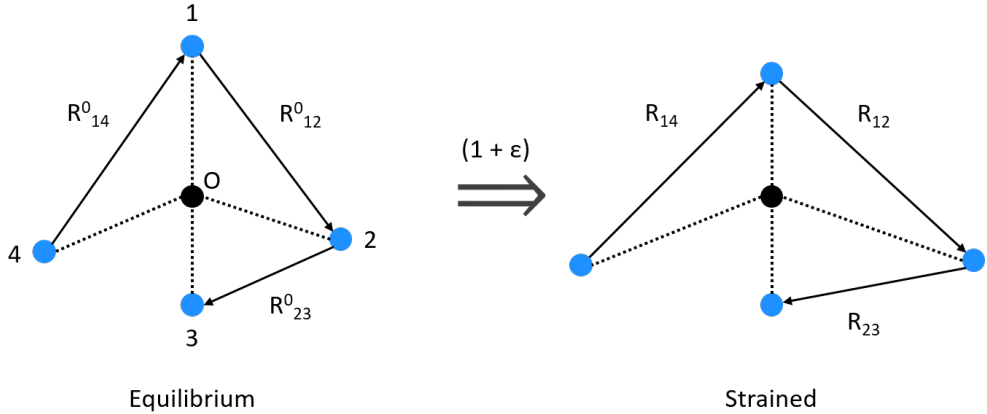

Figure S5: Schematic illustration of vectors at equilibrium and after strain ($\epsilon$) is applied to determine the local strain tensor at the central atom $O$.

9

calculations. From this, we find that the strain tensor for the "close" configuration is diagonal, i.e. $\epsilon_{i,j} = 0$ for $i \neq j$, with $\epsilon_{xx} = 0.28$, $\epsilon_{yy} = 0.27$ and $\epsilon_{zz} = 0.16$. We observe a stronger asymmetry between in-plane and out-of-plane strain relaxation, which as discussed below, is also important for built-in field effects as the bonds in the $c$-plane and along the $c$-axis are distorted differently.

Equipped with information about the strain tensor, $\epsilon$, we utilize $\mathbf{k} \cdot \mathbf{p}$ theory in conjunction with Pikus-Bir strain corrections to gain insight into strain induced energetic shifts of valence states in BN [25]. Following Ref. [25] and the fact that the valence band maximum in BN and GaN is located at the Γ-point ($\mathbf{k} = \mathbf{0}$), we determine the energetic shifts of $p_x$-, $p_y$- and $p_z$-like basis states via:

$$H_{str} = \begin{pmatrix} S_x & 0 & 0 \\ 0 & S_y & 0 \\ 0 & 0 & S_z \end{pmatrix}, \tag{2}$$

$$S_x = (D_2 + D_4)(\epsilon_{xx} + \epsilon_{yy}) + D_5(\epsilon_{xx} - \epsilon_{yy}) + (D_1 + D_3)\epsilon_{zz}, \tag{3}$$

$$S_y = (D_2 + D_4)(\epsilon_{xx} + \epsilon_{yy}) - D_5(\epsilon_{xx} - \epsilon_{yy}) + (D_1 + D_3)\epsilon_{zz}, \tag{4}$$

and

$$S_z = D_2(\epsilon_{xx} + \epsilon_{yy}) + D_1\epsilon_{zz}, \tag{5}$$

where $D_i$ are the valence band deformation potentials; the required $D_i$ values are taken from recent HSE-DFT calculations for wz BN [11] and the strain tensor components $\epsilon_{ij}$ from our analysis above. Thus, for our (B,Ga)N alloy with 4 B atoms in the "close" configuration we find that $p_x$- and $p_y$-like states in BN would be shifted by 0.4 eV ($S_x = S_y = 0.4$ eV) which would not be large enough to overcome the band offset of $-1$ eV, by which the valence band edge in BN "sits" below the valence band edge in GaN. However, for $p_z$-like states we find a shift exceeding this band offset value ($S_z = 1.5$ eV). We stress that this is a simplified approach as the surrounding Ga atoms in (B,Ga)N will also be subject to local strain effects which can affect their band edge values. However, to gain first insight into the impact of local strain effects on the valence band edge states, the above should be sufficient.

## S5: Polarization Vector Field

Given that the bonds in wz III-N are highly polar, these systems in general exhibit very large built-in polarization vector fields related to spontaneous and piezoelectric contributions [26–28]. In following we present a simple approach that allows us to gain first insights into the impact of local strain effects on the local polarization fields. Our model will focus only on the ionic contributions to the polarization vector field and exclude the electronic contributions. For our purposes, again such a simplified treatment is sufficient as the ionic contributions are already significant, as demonstrated below.

In general, the polarization vector field $\mathbf{P}$ can be expressed as:

$$\mathbf{P} = \frac{\boldsymbol{\mu}_0}{V}, \tag{6}$$

where $\boldsymbol{\mu}_0$ is the (net) dipole moment per tetrahedron and $V$ is the volume. The dipole moment $\boldsymbol{\mu}_0$ can be obtained via

$$\boldsymbol{\mu}_0 = -\frac{e^*}{4}(\mathbf{b}_1 + \mathbf{b}_2 + \mathbf{b}_3 + \mathbf{b}_4), \tag{7}$$

where $e^*$ is the born effective charge and $\mathbf{b}_i$ are the bond vectors as labelled in Fig. S6.

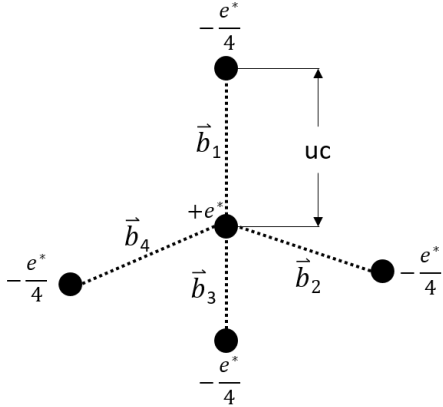

Figure S6: Schematic illustration of bond vectors in a tetrathedron that are use in our model to estimate the local polarization.

As discussed above and when only accounting for *ionic* contribution to the polarization $\mathbf{P}$, in an *ideal* wz structure no polarization is expected. This holds true when applying Eq. (7) to an unstrained tetrahedron. Here, $\mathbf{b}_1 = (0, 0, b_1)^T$, for example, would be pointing perfectly along the $c$-axis; "T" denotes the transpose of the vector. In an unstrained system, $b_1 = uc$, where $u$ is the internal parameter and $c$ is the lattice constant of the crystal structure; for an *ideal* wz system $u = \frac{3}{8}$ [29]. Moreover, when employing Eq. (7), one finds that in the ideal wz case, $x$- and $y$ components of the bond vectors $\mathbf{b}_2$, $\mathbf{b}_3$ and $\mathbf{b}_4$ cancel and only the $z$-component is non zero. In this case, $\boldsymbol{\mu}_0^{\text{ideal}} = e^*(u - \frac{3}{8})c\,\mathbf{e}_z$, where $\mathbf{e}_z$ is the unit vector along the $c$-axis ($z$-axis). Given that in an ideal wz system $u = \frac{3}{8}$, $\boldsymbol{\mu}_0^{\text{ideal}} = \mathbf{0}$, and thus there is no built-in polarization.

We note that for all (B,Ga)N alloy calculations, such as the 4 B atom configuration discussed in the previous sec-

tion, where the BN tetrahedron is distorted, the general expression, Eq. (7), is employed to estimate the polarization $\mathbf{P}$, Eq. (6). Using the atomic positions determined from our DFT calculations, for the 4 B "close" configuration a dipole moment of $\boldsymbol{\mu}_0^{\text{BN,close}} = \left(0.18\,\text{C\AA}\right)\mathbf{e}_z$ was found. In bulk GaN, we find a value of $\boldsymbol{\mu}_0^{\text{GaN, bulk}} = \left(-0.02\,\text{C\AA}\right)\mathbf{e}_z$, as even in a bulk system the bond lengths deviate from the ideal wz crystal structure. The required Born effective charges $e^*$ have been taken from Refs. [30] (wz BN) and [26] (wz GaN). By calculating the volume $V$ of a tetrahedron, using $V = \frac{1}{3}Ah$, where $A$ is the area of the base and $h$ is the height of the tetrahedron, we then determine the polarization $\mathbf{P}$ via Eq. (6). The tetrahedron of the 4 B "close" configuration has a volume of 4.01 $\text{\AA}^3$, and the total polarization is $\mathbf{P}^{\text{BN,close}} = (0, 0, P_z^{\text{BN,close}})^T$ with $P_z^{\text{BN,close}} = 0.045$ C$\text{\AA}^{-2}$. The volume of a relaxed GaN tetrahedron is 4.16 $\text{\AA}^3$, which results in a polarization $\mathbf{P}^{\text{GaN,bulk}} = (0, 0, P_z^{\text{GaN,bulk}})^T$ with $P_z^{\text{GaN,bulk}} = -0.005$ C$\text{\AA}^{-2}$. Any discontinuity in the polarization vector field of the structure can now lead to local (bound) polarization charges, which thus can results in locally strong carrier confinement and in carrier localization. As the local BN tetrahedron has a vastly different polarization in comparison to GaN, this therefore can contribute to carrier localization effects.



\end{document}